\documentclass[aip,amsmath,amssymb,reprint]{revtex4-1}

\usepackage{graphicx}
\usepackage{dcolumn}
\usepackage{bm}

\usepackage[utf8]{inputenc}
\usepackage[T1]{fontenc}
\usepackage{mathptmx}

\begin{document}

\preprint{AIP/123-QED}

\title{A data-driven approach to determine dipole moments of diatomic molecules}

\author{Xiangyue Liu}

\affiliation{Fritz-Haber-Institut der Max-Planck-Gesellschaft, Faradayweg 4-6, 14195 Berlin, Germany }

\author{Gerard Meijer}

\affiliation{Fritz-Haber-Institut der Max-Planck-Gesellschaft, Faradayweg 4-6, 14195 Berlin, Germany }

\author{Jes\'{u}s P\'{e}rez-R\'{i}os}

\affiliation{Fritz-Haber-Institut der Max-Planck-Gesellschaft, Faradayweg 4-6, 14195 Berlin, Germany }
\email{jperezri@fhi-berlin.mpg.de}

\date{\today}

\begin{abstract}
We present a data-driven approach for the prediction of the electric dipole moment of diatomic molecules, which is one of the most relevant molecular properties. In particular, we apply Gaussian process regression to a novel dataset to show that dipole moments of diatomic molecules can be learned, and hence predicted, with a relative error $\lesssim 5\%$. The dataset contains the dipole moment of 162 diatomic molecules, the most exhaustive and unbiased dataset of dipole moments up to date. Our findings show that the dipole moment of diatomic molecules depends on atomic properties of the constituents atoms: electron affinity and ionization potential, as well as on (a feature related to) the first derivative of the electronic kinetic energy at the equilibrium distance. 


\end{abstract}

\maketitle

\begin{quotation}

\end{quotation}

\section{Introduction}

The study of relationships between spectroscopic constants is a traditional topic in chemical physics since the pioneering work of Kratzer and Mecke, among others\cite{Kratzer1920,Mecke1925,Morse1929,Badger1934,Clarck1934,Clark1934bis} and is beautifully summarized by Varshini.\cite{Varshni1957,Varshni1958} Recently, we have shown that some spectroscopic constants are universally related,\cite{liu2020universality} i.e., the relationships between them are independent of the nature of the molecular bond. However, the electric dipole moment of a molecule, despite being an essential molecular property, has not been considered in previous studies about relationships between spectroscopic constants. Only recently, there have been some efforts towards the understanding of the dipole moment in terms of molecular spectroscopic constants. As a result, it has been found by Hou and Bernath that the expression for the dipole moment, $d$, taught in elementary chemistry courses

\begin{equation}
\label{eq1}
d=qR_e,
\end{equation}
\noindent
where $q$ is the effective charge and $R_e$ denotes the equilibrium bond length of the molecule, does not capture the proper physics of the dipole moment in many molecules.~\cite{Hou2015,Hou2015bis} They also demonstrated that the dipole moment of some molecules can be predicted from the effective charge (obtained from quantum chemistry calculations) and spectroscopic constants of molecules. 

In the 2000s the big data-driven science paradigm emerged in the scientific community.\cite{Schleder_2019} In this new paradigm, machine learning techniques are among the most prominent tools to assess scientific knowledge. To be precise, adequately formatted data are used to identify unexpected correlations and to predict observables based on patterns and trends of the data. When applied to physics, this novel paradigm lets nature speak up through hidden and intriguing correlations that lead to the formulation of new questions beyond a specific physical model. In particular, in chemical physics, as recently shown, data-driven approaches bring a new perspective to solve some of the most delicate problems of the field.\cite{Dral2020,Noe2020,Jorg2016,Krems2019}

\begin{figure}
    \centering
    \includegraphics[width=0.8\linewidth]{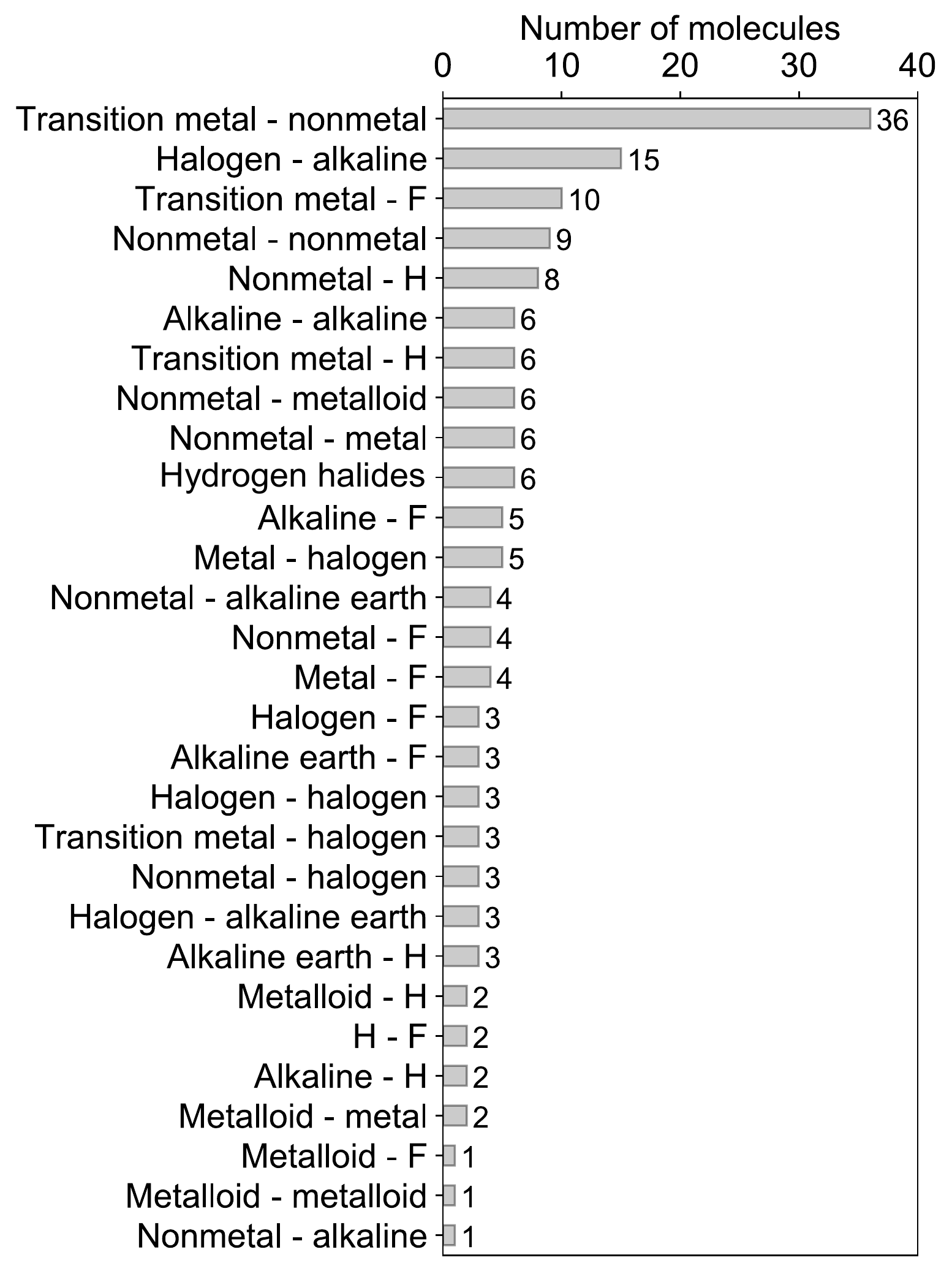}
        \caption{Molecules in the whole dataset classified by the types of their constituent atoms.}
        \label{Fig:mol_class}
\end{figure}

In this paper, we present a data-driven approach to dipole moments of diatomic molecules and its relationship with spectroscopic constants. We show that, after compiling the most exhaustive list of dipole moments for diatomics up to date (to the best of our knowledge) into a dataset, it is possible to learn the dipole moment of diatomic molecules based upon atomic and molecular properties with an error $\lesssim 5\%$. The number of molecules in our dataset, classified by the type of the constituent atoms, is given in Fig.~\ref{Fig:mol_class}. Our results reveal that it is not possible to predict the dipole moment of a molecule solely from atomic properties, although this is feasible for the spectroscopic constants,\cite{liu2020universality} but that it is necessary to include molecular features. The molecular spectroscopic constants are needed in a combination that describes the force on the electrons at the equilibrium distance, i.e. in a combination that has the same functional dependence as the first derivative of the electronic kinetic energy at the equilibrium distance. 

\section{An overview on the nature of the electric dipole moment of molecules}

The study of the nature of the electric dipole moment of molecules is a traditional topic in quantum chemistry that has fascinated the chemical physics community for almost a century by now. The first explanation of the nature of the electric dipole moment of molecules is due to Pauling in the 1930s\cite{PaulingBook}. In particular, after studying hydrogen halide molecules, Pauling proposed that the dipole moment of a molecule is correlated with the relevance of the ionic structure with respect to the covalent one at the equilibrium bond length of the molecule. In this model, the dipole moment is a consequence of the charge transfer between the atoms within the molecule. Therefore, the larger the charge transfer, the bigger the dipole moment is. The charge transfer is quantized by the ionic character (IC), which is given by  

\begin{equation}
\label{eq2}
    \text{IC}=\frac{d}{eR_e},
\end{equation}

\noindent
where $e$ is the electron charge. Comparing Eqs.~(\ref{eq2}) and (\ref{eq1}), it is clear that the ionic character is equivalent to the effective charge, $q$, placed at the center of each of the atoms forming the molecule, as prescribed by Eq.~(\ref{eq1}). However, Pauling's model does not predict 100$\%$ of ionic character for molecules that are fully ionic, like the alkali metal halides. Despite the slight inaccuracy of Pauling's model in predicting dipole moments, it is worth emphasizing that Pauling realized that the dipole moment of a molecule must be related to other molecular properties through the molecular bond.

The next step towards understanding the electric dipole moment was the introduction of a new concept: the homopolar dipole moment, $d_{h}$, by Mulliken. In particular, Mulliken realized that because the atomic orbitals are different in size, the overlap between those leads to a charge displacement with respect to the midpoint of the equilibrium bond length, which affects the electric dipole moment of the molecule~\cite{Mulliken1935}. Furthermore, Mulliken noticed that the asymmetry in the charge distribution of hybrid orbitals causes the so-called atomic dipole moment, $d_a$. The models of Mulliken and Pauling were summarized and further developed by Coulson~\cite{CoulsonBook}, who proposed the ultimate expression for the dipole moment of a diatomic molecule as 

\begin{equation}
\label{eq3}
d=eR_e+d_{a}+d_{h}+d_{p},    
\end{equation}

\noindent
where $d_p$ is the contribution due to the polarization of the atomic orbitals to the dipole moment of the molecule. One has to realise that Eq.~(\ref{eq3}), although being more precise than Eq.~(\ref{eq1}), requires the input from quantum chemistry calculations. For a summary on the Pauling and Mulliken models, we recommend the comprehensive review of Klessinger.\cite{Klessinger1970}

The models of Pauling and Mulliken have been accepted by the physical chemistry community and taught in elementary chemistry courses for a long time, despite the fact that neither one of those is fully satisfactory. Recently, Hou and Bernath\cite{Hou2015,Hou2015bis}, after studying the experimentally determined dipole moments of an extensive group of molecules and using quantum chemistry calculations, have suggested that the electric dipole moment of a molecule should be given as  

\begin{equation}
\label{eq4}
d=qR_d    
\end{equation}

\noindent
where $q$ is the effective charge and $R_d$ is an effective length that depends on fundamental spectroscopic constants of the molecule with $R_d< R_e$. Both Eq.~(\ref{eq4}) and Eq.~(\ref{eq3}) rely on the input of quantum chemistry calculations, in particular on the results from a natural bond orbital analysis. Therefore, the electric dipole moment of diatomic molecules still lacks a satisfactory and accurate explanation in terms of fundamental spectroscopic constants. 

\section{Machine learning model }

\subsection{Gaussian process regression}
Finding relationships of the dipole moment with spectroscopic constants can be viewed as a regression problem, where the goal is to learn the mapping from the input atomic and molecular features $\mathbf{x}$ onto the target property, $y$, which in this case is the electric dipole moment, by a function $y=f(\mathbf{x})$. In the present work, we use Gaussian process regression (GPR) to approximate the function $f(\mathbf{x})$. As a non-parametric probabilistic method, GPR does not presume a functional form of $f(\mathbf{x})$ before observing the data. Instead, it infers a Gaussian distribution of functions over function space by a Gaussian process \cite{williams2006gaussian, MATLAB}

\begin{equation}
    f(\mathbf{x}) \sim \text{GP}(m(\mathbf{x}), k(\mathbf{x}, \mathbf{x}')),
\end{equation}

\noindent
determined by a mean function, $m(\mathbf{x})$, and a kernel (covariance) function, $k(\mathbf{x}, \mathbf{x}')$. The prior, $p(f|\mathbf{x})$, spanning in the function space, after exposed to the observations, is constrained into a posterior, $p(f|\mathbf{x}, y)$, based on the Bayes theorem. The predictions, $y^*$, can then be made for new input atomic and molecular features, $\mathbf{x}$, through the posterior.

The kernel function, $k(\mathbf{x}, \mathbf{x}')$, captures the smoothness of the response and intrinsically encodes the behaviour of the model acting on the input. The kernel functions can be chosen by presuming the behaviour of the response to the input feature by observing the data. Its functional form and the possible hyperparameters can also be determined by a cross-validation (CV)~\cite{raschka2018model}.

\subsection{Model evaluation}

In learning the dipole moments, the dataset is divided into training and test sets. As a data-driven approach, GPR learns the relationship between the input features and dipole moments by observing the training set, while the predictive performance of the GPR models is examined with the test set. In this work, 20 molecules are used in the test set, while the rest are used in the training set. For the training/test splitting, the dataset is first stratified into 20 strata based on the dipole moments' true values. A Monte Carlo (MC) approach is then performed to select the $20$ test data from the dataset randomly. In each MC step, a GPR model is trained based on the training set with $5$-fold cross-validation. The generalization performance of the model is then evaluated with the test set. In the end, the mean and standard deviation (STD) of the test-set errors are reported in this work, obtained from $1000$ MC training/test splittings. Details about this MC approach will be discussed elsewhere.~\cite{MC}

The performance evaluation of the GPR models is carried out through three different estimators:

\begin{itemize}
    \item The mean absolute error (MAE) defined as

\begin{equation}
    \text{MAE} = \frac{1}{N} \sum_{i=1}^{N}|y_i - y^*_i|,
\end{equation}

\noindent
where $y_i$ are the true values of dipole moments, $y^*_i$ are the predictions, and $N$ is the number of observations in the dataset. 

\item The root mean square error (RMSE), which reads as
\begin{equation}
    \text{RMSE}= \sqrt{\frac{1}{N}\sum_{i=1}^{N}{(y_i - y^*_i)}^2}.
\end{equation}

\item The normalized error, $r_E$, defined as the ratio of the RMSE to the range of the data 
\begin{equation}
r_E = \frac{\text{RMSE}}{y_{max}-y_{min}}.
\end{equation}

\end{itemize}

\section{The dataset}

\begin{figure}[h]
    \centering
    \includegraphics[width=0.8\linewidth]{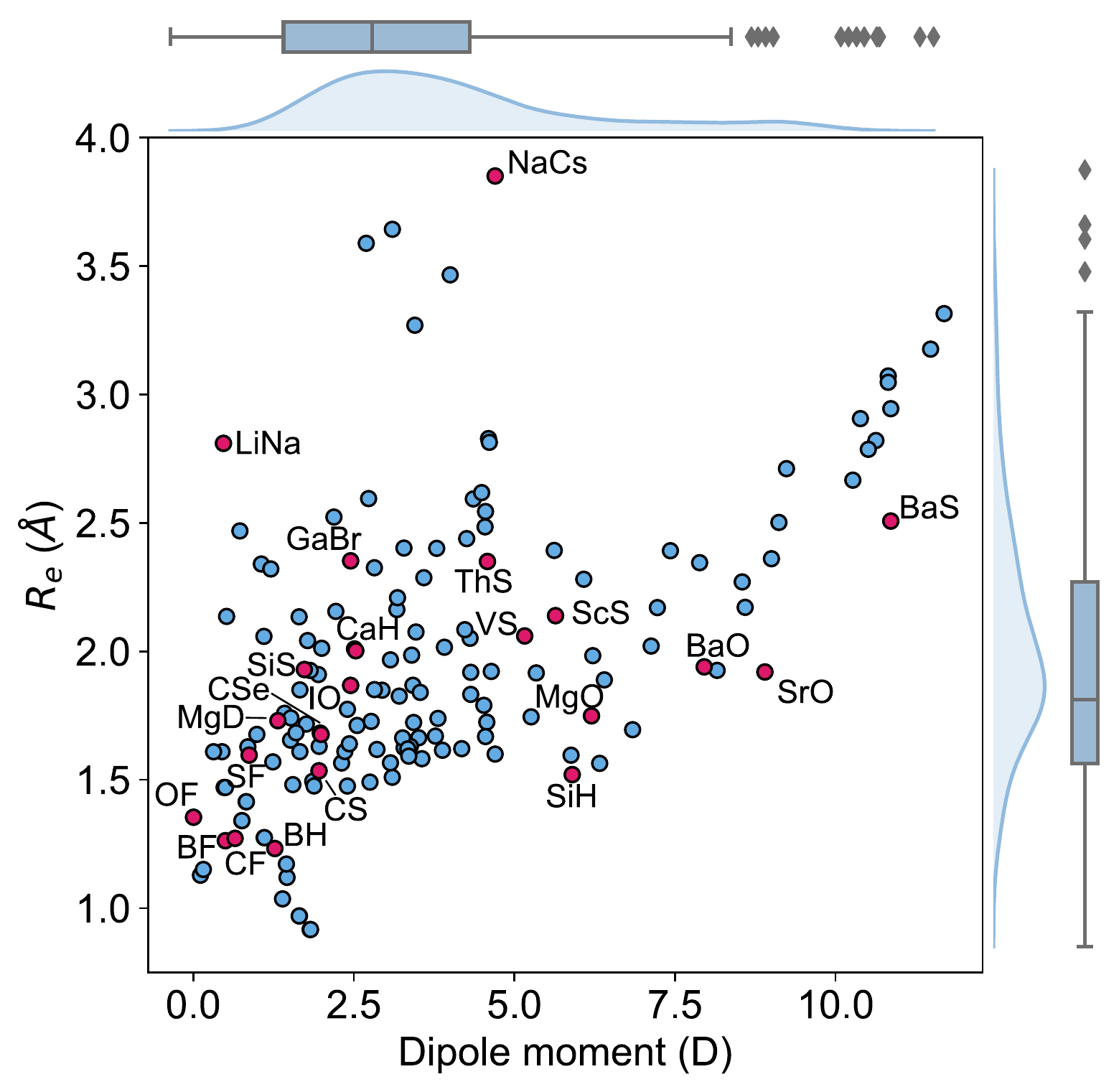}
        \caption{The equilibrium bond length $R_e$ versus the electric dipole moment of the molecules in the dataset. The blue filled circles are the molecules that can be learned by the GPR model in this work. The red filled circles indicate the molecules that can hardly be described by the GPR model in this work. These molecules are labeled by their chemical formula. The density in the right part and upper part of the figure shows the kernel density distribution of $R_e$ and dipole moments, respectively. The box plot shows the minimum, the maximum, the sample median, and the first and third quarterlies of $R_e$ (right) and dipole moments (top).}
        \label{Fig:Re_vs_dipole}
\end{figure}

The dataset employed in this work consists of ground-state dipole moments of $162$ polar diatomic molecules, $139$ of which have both information on the equilibrium bond length, $R_e$, and the harmonic vibrational frequency, $\omega_e$. The dataset is presented in Table~\ref{tab:data_all_references} of the Appendix and it constitutes the most extensive dataset for dipole moments of diatomic molecules that we are aware of. Nevertheless, for more efficient scrutiny of our dataset's generality, we show in Fig.~\ref{Fig:Re_vs_dipole} the equilibrium bond length, $R_e$, versus the electric dipole moment of diatomic molecules. The density plots and the box plots show the distribution of $R_e$ (right) and dipole moment, $d$, (top), respectively. The equilibrium bond length of the molecules is distributed between $0.9$ and $3.9$ \AA ~with a median of around $1.5$ \AA, although most of the molecules show an equilibrium bond length between $1.2$ and $3.2$~\AA. The dipole moment values in the dataset range from $0.0043$~D to $11.69$~D with a median of around $2.45$~D, which shows the large variety of molecules included in the dataset. 

The dataset can also be categorized in terms of the type of atoms constituting the molecules, as it is shown in Fig.~\ref{Fig:mol_class}. In this figure, it is noticed that most of the molecules in the dataset present a highly ionic bond resulting from a transition metal and a nonmetal atom. The second most prominent group of molecules contains a halogen atom and an alkaline atom, which shows an ionic bond. The rest of the molecules exhibit a bond from partially ionic to highly ionic, which shows the diversity of the dataset.

\begin{table*}
\caption{\label{table:GPR_predictions}GPR Predictions on the ground-state dipole moments. $g_i$, $p_i$, EA$_i$, IP$_i$, $\chi_i$, $\alpha_i$ are groups, periods, electron affinity, ionic potential, electronegativity and polarizability of the atom $i$, respectively. $\mu$ is the reduced mass of a molecule. For these results we employ $118$ from the dataset out of the $139$ molecules having values for both $R_e$ and $\omega_e$. }
\begin{ruledtabular}
\begin{tabular}{lccr}

Feature & Test RMSE (D) & Test MAE (D) & Test $r_E$ ($\%$)\\
\hline
(EA$_1$,EA$_2$,IP$_1$,IP$_2$, $\sqrt{\mu R_e \omega_e^2}$) & $0.56\pm0.02$ & $0.43\pm0.0004$   & $4.8\pm0.1$\\

($\chi$$_1$, $\chi$$_2$, $\sqrt{\mu R_e \omega_e^2}$) & $0.70\pm0.05$ & $0.52\pm0.03$   & $6.0\pm0.4$\\

(EA$_1$,EA$_2$,IP$_1$,IP$_2$, $\chi$$_1$, $\chi$$_2$) & $0.86\pm0.006$ & $0.65\pm0.02$   & $7.4\pm0.05$\\

(EA$_1$,EA$_2$,IP$_1$,IP$_2$) & $0.97\pm0.05$ & $0.74\pm0.05$   & $8.3\pm0.4$\\
(EA$_1$,EA$_2$,IP$_1$,IP$_2$, $R_e$) & $1.04\pm0.02$ & $0.81\pm0.04$   & $9.1\pm0.2$\\

($\chi1$, $\chi_2$, $\alpha_1$, $\alpha_2$) & $1.29\pm0.004$ & $1.01\pm0.007$   & $11.2\pm0.04$\\

($\chi$$_1$, $\chi$$_2$) &
$1.35\pm0.002$ & $1.05\pm0.009$   & $11.7\pm0.01$\\

($\sqrt{|\chi_1-\chi_2|}$, $\bar{\alpha}$, $D_0^{-1}$) & $1.21\pm0.03$ & $0.96\pm0.03$   & $10.5\pm0.3$\\

($p_1$,$p_2$,$g_1$,$g_2$, $R_e$) & $1.25\pm0.02$ & $0.94\pm0.04$   & $10.8\pm0.1$\\
\end{tabular}
\end{ruledtabular}
\end{table*}

\section{Results and discussion}

We have used a GPR approach to learn the diatomic molecules' dipole moment employing features coming from different atomic and molecular properties. The atomic properties considered are the electron affinity (EA), ionic potential (IP), electronegativity ($\chi$) and polarizability ($\alpha$) whereas the molecular properties are the reduced mass, $\mu$, equilibrium bond length, $R_e$, and the harmonic vibrational frequency, $\omega_e$. The atomic properties employed are related to the intrinsic chemical nature of the dipole moment due to the polarity of a molecular orbital in the molecular-orbital bond theory or to the ionic character of the molecular bond within the valence-bond theory.~\cite{CoulsonBook} The GPR performance for different features is summarized in Table~\ref{table:GPR_predictions}, where we employ $118$ out of the $139$ molecules from the dataset having values for both $R_e$ and $\omega_e$.

\begin{figure}
    \centering
    \includegraphics[width=0.8\linewidth]{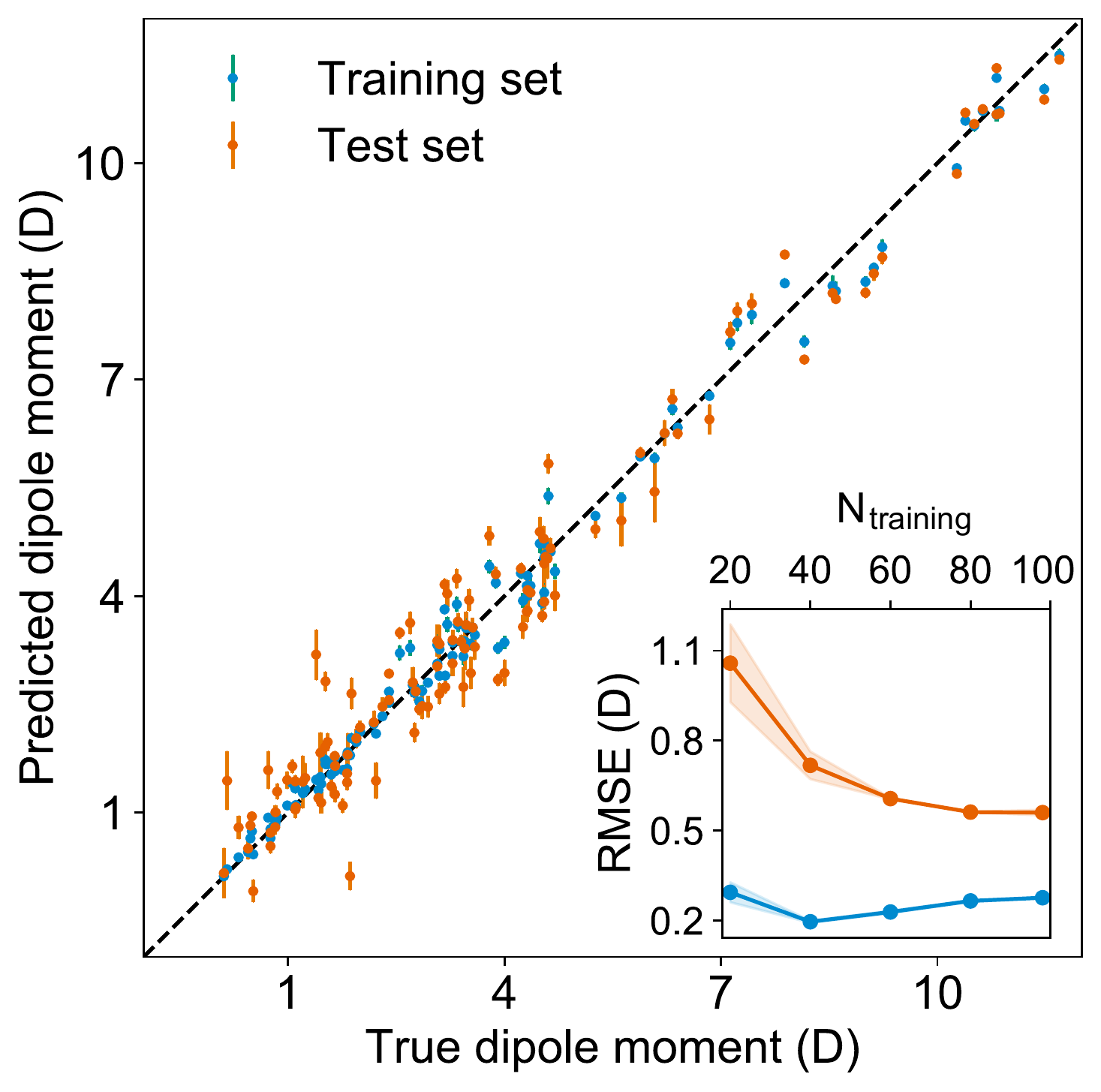}
        \caption{The GPR predictions of the ground-state dipole moments. The values shown in this figure are the average of predictions from 1000 MC sampled training/test splittings \cite{MC}. The test set contains $20$ molecules, while the training set contains $98$ molecules. The mean and standard derivation of the predictions are shown for each molecule when they are used as training data (shown in blue) and test data (shown in orange). The inset shows the learning curve, which shows the training and test RMSE of the model with respect to the number of training data. The shade in the learning curve shows the variance of training/test RMSE, obtained for each point from a MC approach of $500$ training/test splittings.}
        \label{Fig:GPR_dipole_pred_vs_true}
\end{figure}

After using different combinations of atomic and molecular properties, we find that the dipole moment of a diatomic molecule can be best learned by a GPR model using (EA$_1$,EA$_2$,IP$_1$,IP$_2$, $\sqrt{\mu R_e \omega_e^2}$) as the input features. The performance of this model is shown in Fig.\ref{Fig:GPR_dipole_pred_vs_true}, The predicted values reproduce the true values very well with a small deviation that leads to a normalized error $r_E < 5\%$ (RMSE$=0.56\pm0.02$ D). We have also computed the learning curve of the cited GPR model, which gives an intuitive idea about the model's learning and generalization performance concerning the size of the training set. The results are shown in the inset of Fig.\ref{Fig:GPR_dipole_pred_vs_true}. The training RMSE and test RMSE are shown as a function of the number of training data points. The learning curve's shade shows the variance of training/test RMSE, obtained for each point from a MC approach of $500$ training/test splittings. The mean test error decreases with increasing training data. In particular, with $80$ training data, the learning curve is almost converged, suggesting that this model can not benefit from more data of the same dataset. The error's variance shows the ability of the model to be employed in different subgroups of molecules. In this case, the variance of test RMSE becomes smaller as the number of training data increases and converges to $<0.02$ D with $60$ training data. 

In previous work, we have shown that $R_e$, $\omega_e$, and the binding energy of a diatomic molecule can be learned through groups and periods of the constituent atoms as features\cite{liu2020universality}. However, the same features dramatically fail in learning the dipole moment. In particular, we find that the test errors are RMSE$=1.25\pm0.02$ D and $r_E=10.8\pm0.1 \%$, respectively. In our view, this is an indication of the more intricate nature of the dipole moment compared to the spectroscopic constants of diatomic molecules.

In Ref.~\cite{pototschnig2016electric} it is shown that the dipole moment of diatomic alkali–alkaline earth molecules can be empirically calculated from the difference in the electronegativity of the constituent atoms $\sqrt{|\chi_1-\chi_2|}$, the mean atomic polarizabilities $\bar{\alpha}=(\alpha_1+\alpha_2)/2$ and the dissociation energy $D_e$. We have generalized this idea trough a GPR model by using ($\sqrt{|\chi_1-\chi_2|}$, $\bar{\alpha}$, $D_0^{-1}$) as features and applied it to the present dataset, despite the fact that alkaline earth-alkaline molecules are absent in the dataset. We have used the binding energy, $D_0$, instead of the dissociation energy, as the former is tabulated more frequently. As a result, the normalized error is $r_E = 10.5\pm0.3 \%$, which indicates that some of the physics behind the dipole moment function of alkali-alkaline earth molecules is applicable to any other molecule. This is an unexpected result that shows the underlying universality of the physics behind the dipole moment.  

The outstanding performance of (EA$_1$,EA$_2$,IP$_1$,IP$_2$, $\sqrt{\mu R_e \omega_e^2}$) as the set of descriptors implies that the accepted picture in chemistry in which the difference of the electronegativity of the atoms within a molecule establishes the ionic character of the molecular bond~\cite{CoulsonBook,Hannay1946,PaulingBook} is not sufficient to characterize the dipole moment of a molecule. When using the electron affinity and the atoms' ionization potential as features, the performance improves by $25\%$. However, only if $\sqrt{\mu R_e \omega_e^2}$ is included as a feature, the dipole moment is predicted with a RMSE below 0.7~D. Therefore, we find that it is essential to add $\sqrt{\mu R_e \omega_e^2}$ as a feature in describing the dipole moment of a diatomic molecule. It can be shown that this feature is related to the derivative of the electronic kinetic energy, $T(R)$, at the equilibrium bond length as~\cite{Borkman1968}

\begin{equation}
-\frac{dT(R)}{dR}\bigg\rvert_{R=R_e}=\mu R_e \omega_e^2,    
\end{equation}

\noindent
which represents a force within the molecule. When equating this force to the pure electrostatic force, one obtains $R_d$ and, through  Eq.(\ref{eq4}), it is then possible to define the ionic character as

\begin{equation}
\label{eq:IC_Hou}
{\text IC}= 100\left(d \sqrt{\mu R_e \omega_e^2}\right)^{1/2},
\end{equation}

\noindent
where the value of IC is given in percent. It is seen that IC does not directly depend upon the electronegativity differences of the atoms, contrary to the accepted picture in chemistry. The feature $\sqrt{\mu R_e \omega_e^2}$ was first introduced by Hou and Bernath~\cite{Hou2015,Hou2015bis} as an empirical relationship, and we use this here to define the ionic character of a molecular bond. 

\begin{figure}[h!!!!!]
    \centering
    \includegraphics[width=0.8\linewidth]{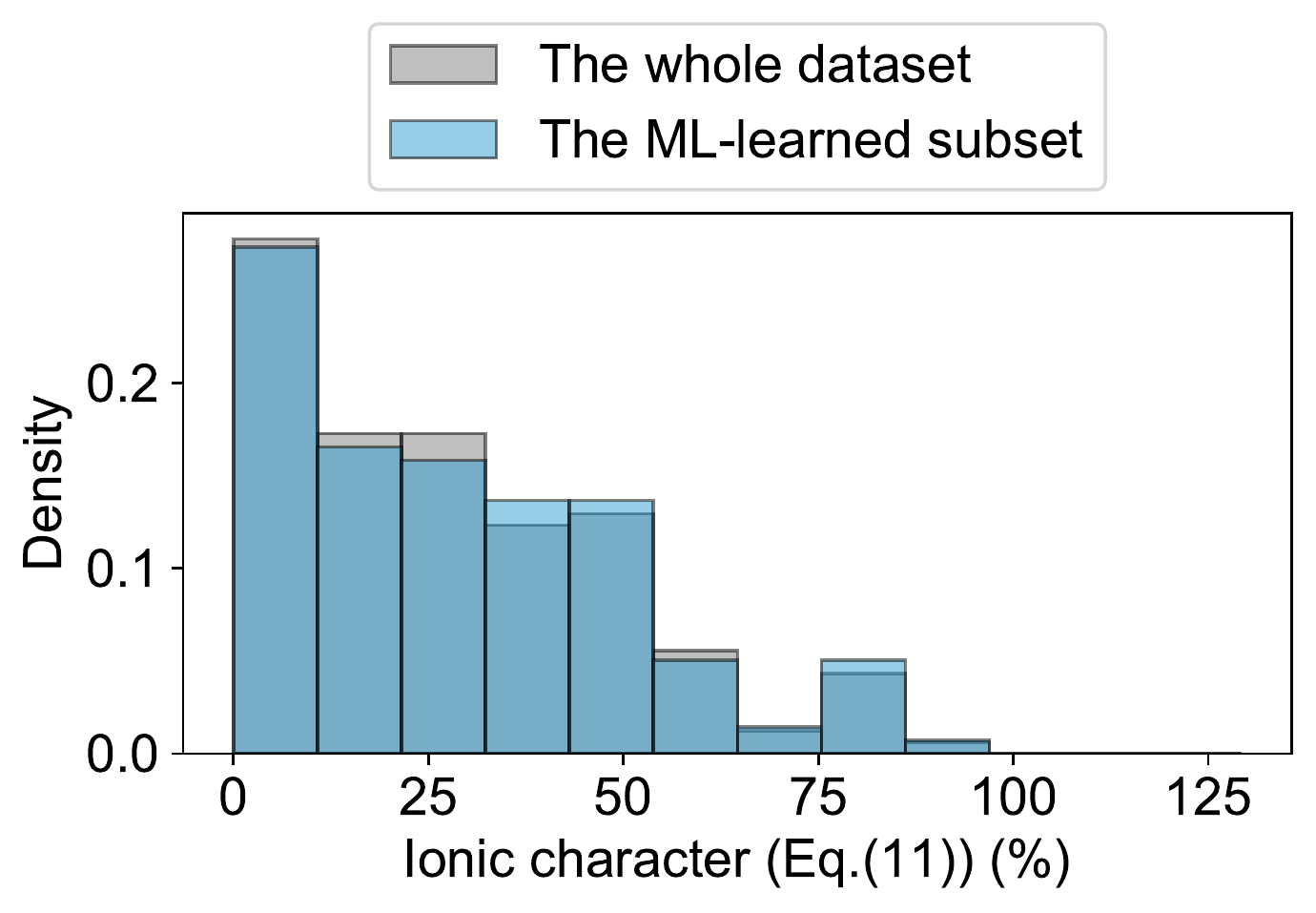}\\
    \includegraphics[width=0.8\linewidth]{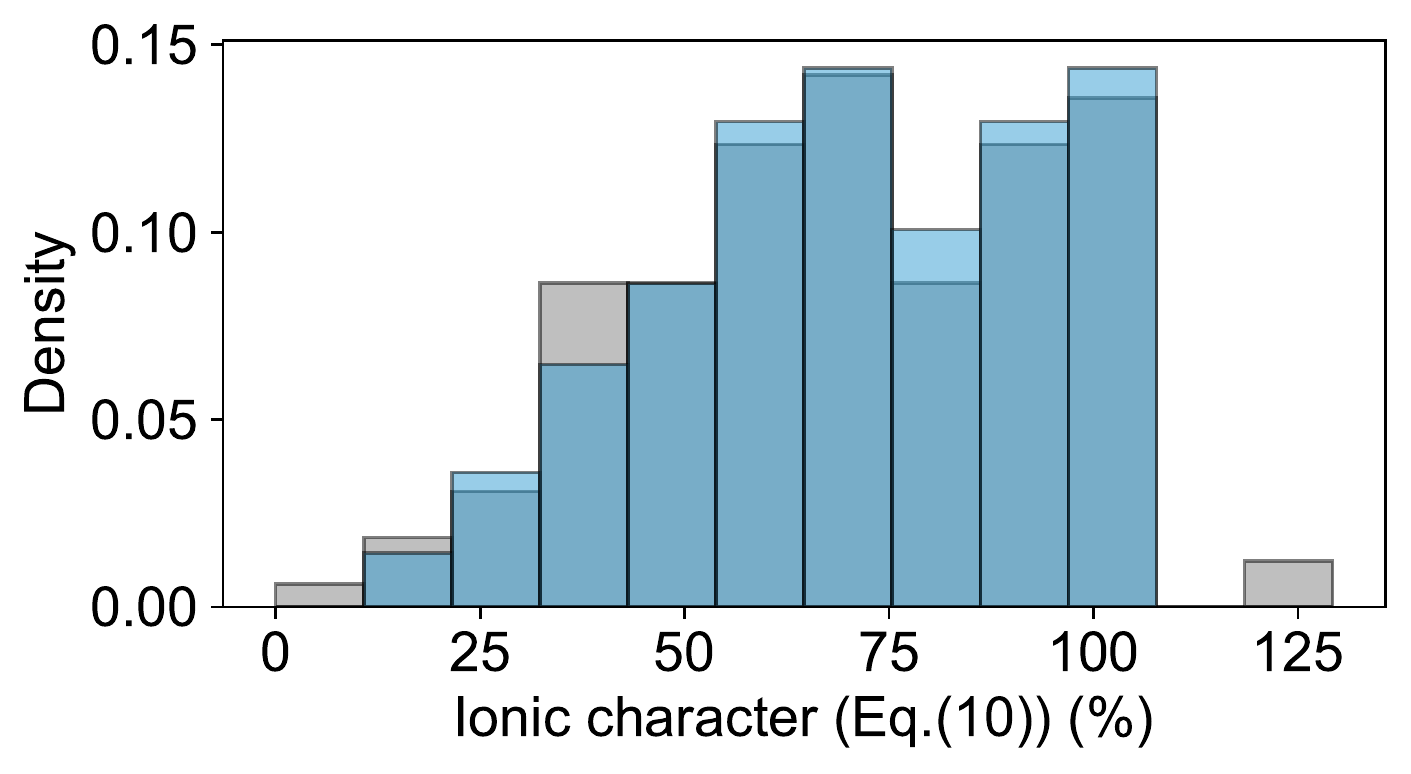}\\
    \includegraphics[width=0.8\linewidth]{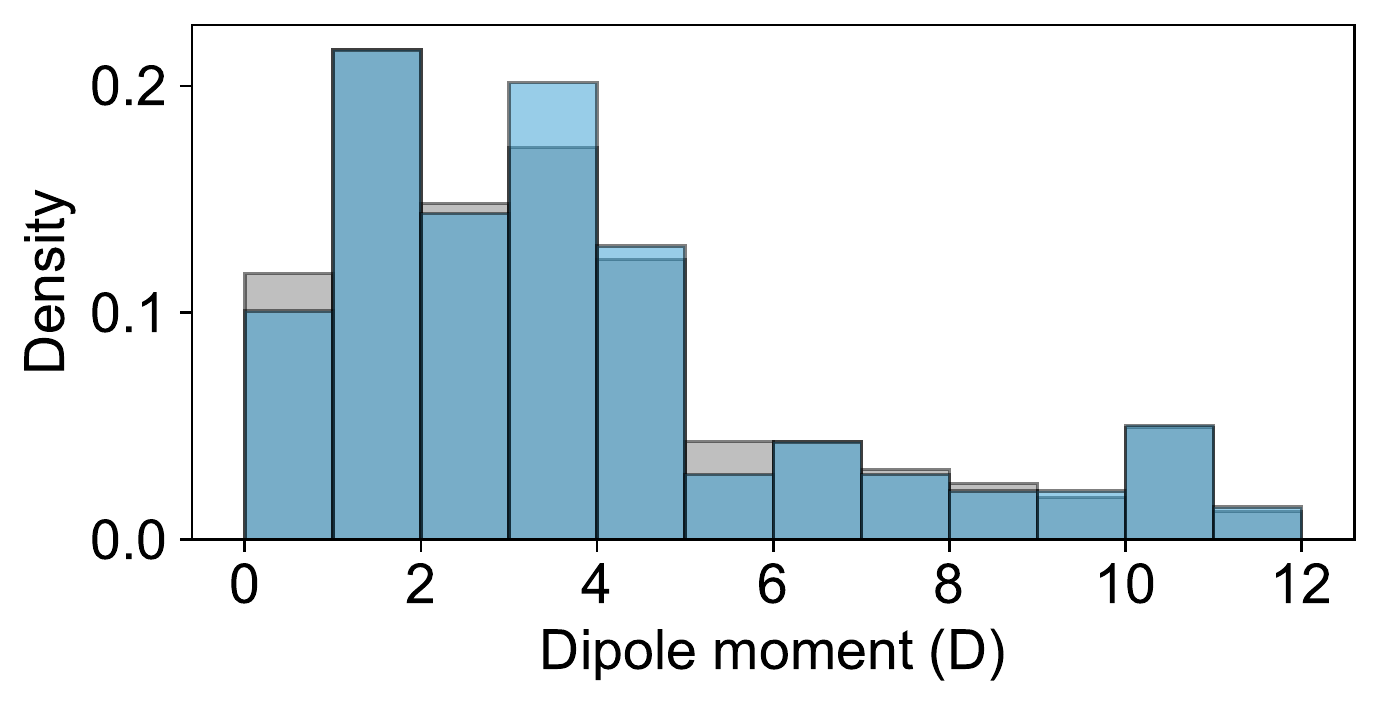}
        \caption{Comparison of the histograms of ionic characters and dipole moments in the whole dataset (shown in grey) and the ML-learned subset of $118$ molecules (shown in blue). Panel (a) and (b) show the ionic characters calculated from Eqs.~(\ref{eq:IC_Hanny}) and (\ref{eq:IC_Hou}), respectively. Panel (c) plots the histogram of the dipole moment of the molecules. It is worth noticing that the dark blue regions appear in regions where the grey and light-blue bars overlap.}
        \label{Fig:Hist_ionic_character_dipole_moment}
\end{figure}

Alternatively, the ionic character can be defined in terms of the electronegativity difference between the two atoms forming a molecule as

\begin{equation}
\label{eq:IC_Hanny}
 \text{IC} = 16|\chi_1 - \chi_2| + 3.5 |\chi_1 - \chi_2|^2, 
\end{equation}

\noindent
following Hannay and Smyth~\cite{Hannay1946}. Surprisingly, Eqs.~\ref{eq:IC_Hou} and \ref{eq:IC_Hanny} lead to different results for the ionic character of the molecules in the database, as shown in Fig.~\ref{Fig:Hist_ionic_character_dipole_moment}, where it is noticed that the distribution of the ionic character following Eq.~\ref{eq:IC_Hanny} appears to the complement to the one obtained from Eq.~\ref{eq:IC_Hou}. Furthermore, the model of Hou and Bernath (Eq.~\ref{eq:IC_Hou}) systematically leads to a larger ionic character than the model of Hannay an Smyth.

The GPR model with (EA$_1$,EA$_2$,IP$_1$,IP$_2$, $\sqrt{\mu R_e \omega_e^2}$) as input features shows several outliers. To see the importance of these outliers we have compared the distribution of the ionic character and dipole moment of the molecules in Fig.~\ref{Fig:Hist_ionic_character_dipole_moment} (shown in grey) with the same magnitudes for the subset of $118$ molecules that can be learned in this work (shown in blue). The ML-learned subset has similar overall distributions of dipole moments and ionic characters in comparison with the whole dataset. Therefore the outliers do not significantly modify the underlying distribution that the molecules follow. 

In Table~\ref{table:GPR_outliers_class}, it is shown a classification of the outliers as a function of its molecular bond and constituent atoms. The effective atomic charges of these molecules are also calculated with a density functional theory (DFT) approach, which is shown in Table~\ref{table:GPR_outliers_charge} utilizing different charge partitioning methods. The calculations are performed with the B3LYP functional\cite{b3lyp} and def2-TZVP basis set~\cite{basiskaupp1991pseudopotential,basisleininger1996accuracy,basisweigend2005balanced}, with the Gaussian 16 package \cite{g16}. We have noticed that for these outliers, the natural bond orbital (NBO) method gives larger effective atomic charges compares to the Mulliken population. Furthermore, all the molecules showing a NBO charge larger than $1.0$ are the ones showing an ionic character in virtue of Eq.\ref{eq:IC_Hou} above 100$\%$. For the outliers within the van der Waals molecules, we find LiNa and NaCs. LiNa has the smallest $R_e$ and dipole moment of the bialkaline molecules in this dataset, while NaCs has the largest $R_e$ and dipole moment.

\begin{table}
\caption{\label{table:GPR_outliers_class} Outliers for learning the electric dipole moment of diatomic molecules. These molecules are labeled in Fig.~\ref{Fig:Re_vs_dipole} and classified with the types of constituent atoms and the molecular bonds.}
\begin{ruledtabular}
\begin{tabular}{cc}
Type of bond	&	Molecule	\\
\hline
Nonmetal-nonmetal	&	IO, CS, SiS, CSe	\\
Nonmetal-F	&	SF, BF, CF, OF	\\
Metal-halogen & GaBr \\
Alkaline earth-nonmetal	&	BaO, SrO, MgO, SrS, BaS	\\
Alkaline earth-H	&	MgD, CaH	\\
Metalloid-H	&	BH, SiH	\\
Transition metal-nonmetal	 &	VS, ScS, ThS	\\
van der Waals	&	LiNa, NaCs	\\
\end{tabular}
\end{ruledtabular}
\end{table}

\begin{figure}
    \centering
    \includegraphics[width=1.0\linewidth]{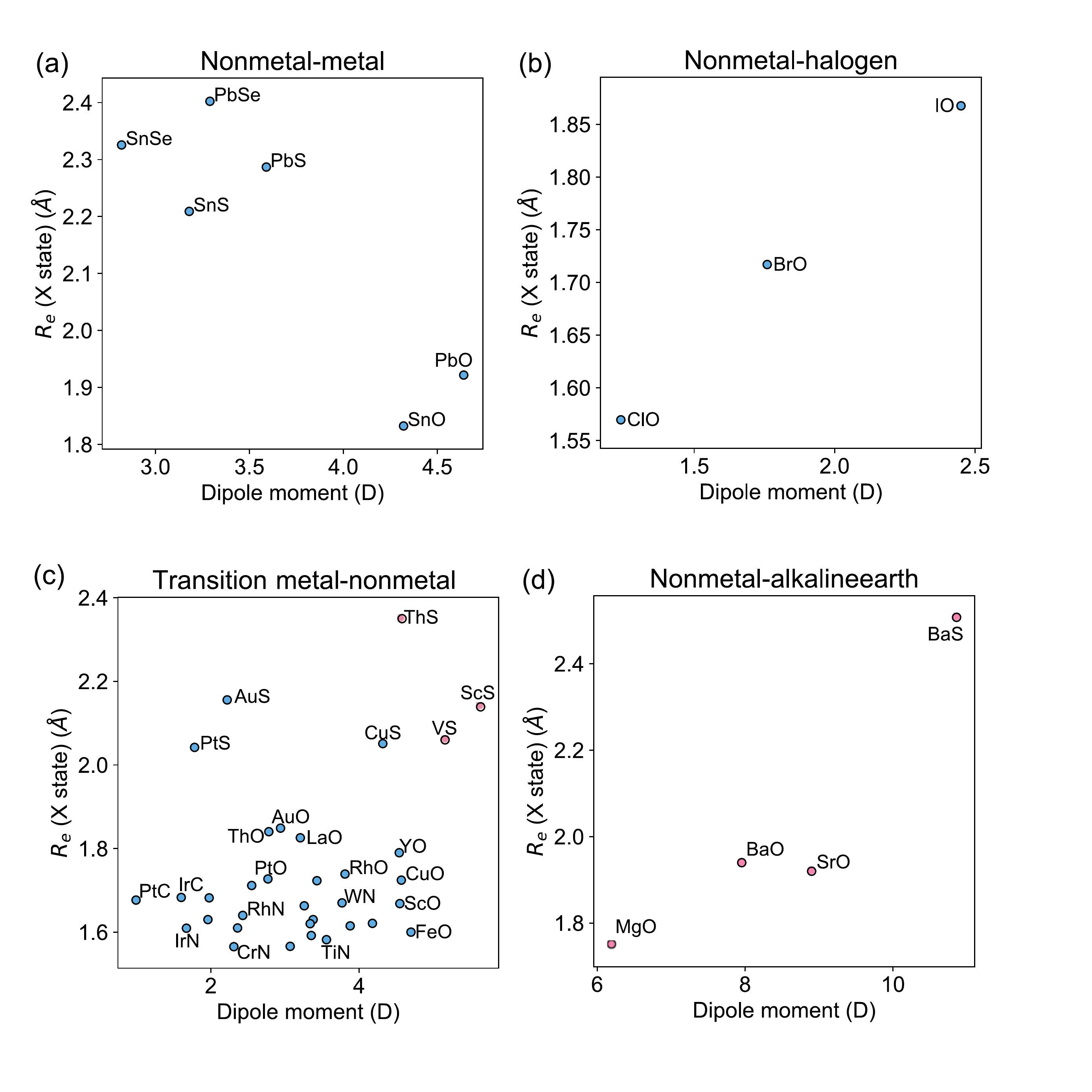}
        \caption{The equilibrium bond lengths $R_e$ as a function of dipole moments, classified by the type of the constituent atoms. The molecules that can be described by the GPR models from (EA$_1$, EA$_2$,IP$_1$,IP$_2$, $\sqrt{\mu R_e \omega_e^2}$) are shown in blue circles, while the outliers are shown in red circles.}
        \label{Fig:Re_vs_dipole_class}
\end{figure}

To understand the effect of different bonding types on the dipole moment, we plot in Fig.~\ref{Fig:Re_vs_dipole_class} the relationships between $R_e$ and dipole moments for different kinds of molecules in the current dataset, where the outliers are shown in red circles. We observe that the relationships between $R_e$ and dipole moments depend on the type of molecule under consideration. As shown in panel (a) of Fig.~\ref{Fig:Re_vs_dipole_class}, $R_e$ and dipole moments show linear relationship for metal-nonmetal molecules, in which the nonmetals atoms are from the same group in the periodic table. Similarly, linear behaviors have also been observed for the group IV/VI diatomic molecules in Ref.~\cite{hoeft1970dipole}. For the oxygen halides shown in panel (b), $R_e$ increases almost linearly with the dipole moment. In panel (c), the molecules containing a transition metal and a nonmetal atom show a different trend of the equilibrium distance as a function of the dipole moment compared with the molecules formed by the main-group metal elements nonmetal atoms in panel (a). Within these molecules, the outliers are the ones with both the largest dipole moments and $R_e$ in panel (c). Interestingly, we find that all the $4$ alkaline earth-nonmetal molecules in the dataset are outliers, as shown in panel (d) of Fig.~\ref{Fig:Re_vs_dipole_class}. Indeed, SrO, BaO and BaS have the largest atomic charges within the molecules in the dataset.

\begin{table}
\caption{\label{table:GPR_outliers_charge} The effective atomic charges of the outliers with different charge partitioning methods, calculated with the B3LYP functional\cite{b3lyp} and def2-TZVP basis set\cite{basiskaupp1991pseudopotential,basisleininger1996accuracy,basisweigend2005balanced} with the Gaussian 16 package \cite{g16}.}
\begin{ruledtabular}
\begin{tabular}{cccc}
Molecule	&	Mulliken	&	Hirschfeld	&	NBO	\\
\hline
MgO	&	0.694	&	0.576	&	1.278	\\
SrO	&	0.871	&	0.714	&	1.496	\\
BaO	&	0.838	&	0.640	&	1.508	\\
BaS	&	0.759	&	0.660	&	1.437	\\
BF	&	0.099	&	0.073	&	0.549	\\
CF	&	0.030	&	0.014	&	0.315	\\
OF	&	0.017	&	0.012	&	0.063	\\
SF	&	0.198	&	0.108	&	0.431	\\
MgD	&	0.187	&	0.241	&	0.657	\\
CaH	&	0.276	&	0.318	&	0.738	\\
BH	&	-0.036	&	0.072	&	0.349	\\
SiH	&	0.048	&	0.122	&	0.349	\\
SiS	&	0.231	&	0.222	&	0.656	\\
CS	&	-0.081	&	-0.087	&	-0.174	\\
SeC	&	0.180	&	0.104	&	0.263	\\
IO	&	0.412	&	0.214	&	0.625	\\
GaBr	&	0.331	&	0.265	&	0.627	\\
ScS	&	0.529	&	0.452	&	0.743	\\
VS	&	0.425	&	0.247	&	0.343	\\
CsNa	&	0.140	&	0.161	&	0.279	\\
NaLi	&	-0.074	&	0.001	&	0.007	\\

\end{tabular}
\end{ruledtabular}
\end{table}

\section{Conclusion}

In summary, we have shown that through a GPR model, the ground state dipole moments of diatomic molecules can be related to spectroscopic constants, namely $R_e$ and $\omega_e$. More specifically, without any quantum chemistry calculation, the dipole moments of $118$ molecules have been predicted with an error $\lesssim 5\%$ by using both atomic features, including electron affinity and ionic potential, and a combination of molecular spectroscopic constants, $\sqrt{\mu R_e \omega_e^2}$. In addition, we find that the difference in the electronegativity of the constituents atoms is not sufficient to describe the dipole moments of the diatomic molecules in stark contrast with what is generally assumed in general chemistry. Therefore, our data-driven approach shows that the nature of the dipole moment is more intricate than other spectroscopic constants, and it is clearly correlated with the very fundamental nature of the chemical bond. Finally, it is worth emphasizing that our findings have been possible thanks to the development of a complete and orthodox dataset.

\begin{acknowledgments}
We thank Dr. Stefan Truppe for reading the manuscript and for useful comments and discussion regarding the nature of the electric dipole moment. 
\end{acknowledgments}

\appendix
\section{Details about GPR}

The kernel function employed in this work, which gives the best CV scores, is the ration quadratic kernel \cite{MATLAB} defined by
\begin{equation}
    k(x_i,x_j|\theta) = \sigma_f^2 \left( 1+ \frac{r^2}{2\alpha\sigma_l^2}\right)^{-a},
\end{equation}
where $\sigma_l$ is the length scale, and $\alpha$ is a scale-mixture parameter, $r$ is the Euclidean distance between $x_i$ and $x_j$ defined as

\begin{equation}
    r = \sqrt{(x_i-x_j)^T(x_i-x_j)}.
\end{equation}

\section{The dataset for dipole moment of diaotmic molecules}

The dataset is summarized in Table \ref{tab:data_all_references}, which consists of dipole moments $\mu_0$ of 162 polar diatmonic molecules, $156$ of which have information about equilibrium bond length $R_e$ while $139$ also have harmonic vibrational frequency $\omega_e$. The references of the dipole moments are also listed in the table.

\begin{table*}
\caption{\label{tab:data_all_references}The dipole moments, $d$, equilibrium bond length, $R_e$, and harmonic vibrational frequency, $\omega_e$, employed in this work. The references to the dipole moments are also listed in the table. $R_e$ and $\omega_e$ are taken from Ref.~\cite{Herzberg}, \cite{Smirnov2008} or the same reference of the dipole moment of the corresponding molecules, except as indicated. }
\begin{ruledtabular}
\begin{tabular}{ccccccccccccccc}

Molecule  &  $d$ (D)  &  $R_e$ (\AA)  &  $\omega_e$ (cm$^{-1}$)  &  Ref. &  Molecule  &  $d$ (D)  &  $R_e$ (\AA)  &  $\omega_e$ (cm$^{-1}$)  &  Ref. &  Molecule  &  $d$ (D)  &  $R_e$ (\AA)  &  $\omega_e$ (cm$^{-1}$)  &  Ref.  \\

\hline

AgBr	&	5.62	&	2.393	&	247.7	&	\cite{haynes2014crc}	&	GeTe	&	1.06	&	2.34	&	323.9	&	\cite{hoeft1970elektrisches}	&	PbSe	&	3.29	&	2.402	&	277.6	&	\cite{hoeft1970elektrisches}	\\
AgCl	&	6.08	&	2.281	&	343.5	&	\cite{haynes2014crc}	&	DBr	&	0.823	&	1.415	&	1884.8	&	\cite{radzig2012reference}	&	PbTe	&	2.73	&	2.595	&	212	&	\cite{hoeft1970elektrisches}	\\
AgF	&	6.22	&	1.983	&	513.5	&	\cite{hoeft1970rotational}	&	HBr	&	0.8272	&	1.414	&	2649	&	\cite{haynes2014crc}	&	PN	&	2.7514	&	1.491	&	1337.2	&	\cite{wyse1972millimeter}	\\
AgH	&	2.86	&	1.618	&	1759.9	&	\cite{sadlej1991mutual}	&	DF	&	1.819	&	0.917	&	2998.2	&	\cite{radzig2012reference}	&	PO	&	1.88	&	1.476	&	1233.3	&	\cite{kanata1988dipole}	\\
AgI	&	4.55	&	2.545	&	206.5	&	\cite{haynes2014crc}	&	HF	&	1.826526	&	0.917	&	4138.3	&	\cite{muenter1970hyperfine}	&	PtC	&	0.99	&	1.677	&	1051.1	&	\cite{steimle1995permanent}	\\
AlF   &	1.515	&	1.654	&	802.3	&	\cite{PhysRevA.100.052513}	&	HfF	&	1.66	&	1.85	&		&	\cite{le2013molecular}	&	PtF	&	3.42	&	1.868	&		&	\cite{qin2012permanent}	\\
AuF	&	4.32	&	1.918	&	539.4 \footnotemark[1]	&	\cite{steimle2013molecular}	&	HfO	&	3.431	&	1.723	&	974.1	&	\cite{suenram1990pulsed}	&	PtN	&	1.977	&	1.682	&		&	\cite{jung1995experimental}	\\
AuO	&	2.94	&	1.849	&	624.59 \footnotemark [2]	&	\cite{zhang2017electric}	&	HI	&	0.448	&	1.609	&	2309	&	\cite{haynes2014crc}	&	PtO	&	2.77	&	1.727	&	851.1	&	\cite{steimle1995permanent}	\\
AuS	&	2.22	&	2.156	&	410.19 \footnotemark[3]	&	\cite{zhang2017electric}	&	IBr	&	0.726	&	2.469	&	268.6	&	\cite{haynes2014crc}	&	PtS	&	1.78	&	2.042	&		&	\cite{steimle1995permanent}	\\
BaF	&	3.17	&	2.163	&	468.9	&	\cite{ernst1986hyperfine}	&	ICl	&	1.207	&	2.321	&	384.3	&	\cite{durand1997spectroscopy}	&	RbBr	&	10.86	&	2.945	&	169.5	&	\cite{story1976dipole}	\\
BaO	&	7.955	&	1.94	&	669.8	&	\cite{wharton1962electric}	&	ID	&	0.316	&	1.609	&	1639.7	&	\cite{burrus1958stark}	&	RbCl	&	10.51	&	2.787	&	228	&	\cite{hebert1968dipole}	\\
BaS	&	10.86	&	2.507	&	379.4	&	\cite{melendres1969radio}	&	IF	&	1.948	&	1.91	&	610.2	&	\cite{haynes2014crc}	&	RbF	&	8.5465	&	2.27	&	376	&	\cite{hebert1968dipole}	\\
BF	&	0.5	&	1.263	&	1402.1	&	\cite{lovas1971microwave}	&	InCl	&	3.79	&	2.401	&	317.4	&	\cite{haynes2014crc}	&	RbI	&	11.48	&	3.177	&	138.5	&	\cite{story1976dipole}	\\
BH	&	1.27	&	1.232	&	2366.9	&	\cite{thomson1969experimental}	&	InF	&	3.4	&	1.985	&	535.4	&	\cite{hoeft1970microwave}	&	ReN	&	1.96	&	0.61	&		&	\cite{steimle2004permanent}	\\
BrCl	&	0.519	&	2.136	&	444.3	&	\cite{haynes2014crc}	&	IO	&	2.45	&	1.868	&	681.5	&	\cite{byfleet1971electric}	&	RhN	&	2.43	&	1.64	&		&	\cite{ma2007molecular}	\\
BrF	&	1.422	&	1.759	&	670.8	&	\cite{haynes2014crc}	&	IrC	&	1.6	&	1.683	&	1060.1	&	\cite{marr1996optical}	&	RhO	&	3.81	&	1.739	&		&	\cite{gengler2007molecular}	\\
BrO	&	1.76	&	1.717	&	778.7	&	\cite{radzig2012reference}	&	IrF	&	2.82	&	1.851	&		&	\cite{zhuang2010electric}	&	RuF	&	5.34	&	1.916	&		&	\cite{steimle2006permanent}	\\
CaBr	&	4.36	&	2.594	&	285.3 \footnotemark[4]	&	\cite{torring1984dipole}	&	IrN	&	1.67	&	1.609	&		&	\cite{marr1996optical}	&	ScO	&	4.55	&	1.668	&	965	&	\cite{shirley1990molecularScO}	\\
CaCl	&	4.257	&	2.439	&	367.5	&	\cite{torring1984dipole}	&	KBr	&	10.6281	&	2.821	&	213	&	\cite{van1967dipole}	&	ScS	&	5.64	&	2.139	&	565.2	&	\cite{steimle1997dipole}	\\
CaD	&	2.51	&	2.01	&		&	\cite{chen2008permanent}	&	KCl	&	10.2688	&	2.667	&	281	&	\cite{hebert1968dipole}	&	SD	&	0.7571	&	1.341	&	1885.5	&	\cite{meerts1975molecular}	\\
CaF	&	3.07	&	1.967	&	581.1	&	\cite{childs1984electric}	&	KF	&	8.59255	&	2.171	&	428	&	\cite{van1967dipole}	&	SeF	&	1.52	&	1.741	&	757	&	\cite{byfleet1971electric}	\\
CaH	&	2.53	&	2.003	&	1298.3	&	\cite{chen2008permanent}	&	KI	&	10.82	&	3.048	&	186.5	&	\cite{story1976dipole}	&	SeD	&	0.48	&	1.47	&	1708	&	\cite{radzig2012reference}	\\
CaI	&	4.5968	&	2.829	&	238.7	&	\cite{ernst1985precise}	&	LaO	&	3.207	&	1.826	&	812.8	&	\cite{suenram1990pulsed}	&	SeH	&	0.5	&	1.47	&	2400	&	\cite{radzig2012reference}	\\
CF	&	0.65	&	1.272	&	1308.1	&	\cite{byfleet1971electric}	&	LiBr	&	7.2262	&	2.17	&	563.2	&	\cite{hebert1964radio}	&	SF	&	0.87	&	1.596	&	837.6	&	\cite{byfleet1971electric}	\\
CH	&	1.46	&	1.12	&	2858.5	&	\cite{haynes2014crc}	&	LiCl	&	7.1289	&	2.021	&	643.3	&	\cite{hebert1968dipole}	&	SH	&	0.758	&	1.341	&	2711.6	&	\cite{meerts1974hyperfine}	\\
ClD	&	1.1033	&	1.275	&	2145.2	&	\cite{kaiser1970dipole}	&	LiF	&	6.32736	&	1.564	&	910.3	&	\cite{hebert1968dipole}	&	SiH	&	5.9	&	1.52	&	2041.8	&	\cite{radzig2012reference}	\\
ClF	&	0.85	&	1.628	&	786.2	&	\cite{fabricant1977molecular}	&	LiH	&	5.882	&	1.596	&	1405.7	&	\cite{wharton1960dipole}	&	SiO	&	3.0982	&	1.51	&	1241.6	&	\cite{raymonda1970electric}	\\
ClH	&	1.1085	&	1.275	&	2990.9	&	\cite{kaiser1970dipole}	&	LiI	&	7.4285	&	2.392	&	498.2	&	\cite{breivogel1965radio}	&	SiS	&	1.73	&	1.73	&	749.6	&	\cite{lovaselektrisches}	\\
ClO	&	1.239	&	1.57	&	853.8	&	\cite{amano1969microwave}	&	LiK	&	3.45	&	3.27	&	207	&	\cite{haynes2014crc}	&	SiSe	&	1.1	&	2.058	&	580	&	\cite{hoeft1970dipole}	\\
CN	&	1.45	&	1.172	&	2068.6	&	\cite{thomson1968experimental}	&	LiNa	&	0.47	&	2.81	&	256.8	&	\cite{dagdigian1971stark}	&	SnO	&	4.32	&	1.833	&	814.6	&	\cite{hoeft1970dipole}	\\
CO	&	0.112	&	1.128	&	2169.8	&	\cite{burrus1958stark}	&	LiO	&	6.84	&	1.695	&	851.5	&	\cite{haynes2014crc}	&	SnS	&	3.18	&	2.209	&	487.3	&	\cite{hoeft1970dipole}	\\
CoF	&	2.82	&		&		&	\cite{wang2009permanent}	&	LiRb	&	4.0	&	3.466	&	195.2	&	\cite{haynes2014crc}	&	SnSe	&	2.82	&	2.326	&	331.2	&	\cite{hoeft1970dipole}	\\
CoH	&	1.88	&		&		&	\cite{wang2009permanent}	&	MgD	&	1.318	&	1.73	&	1077.9	&	\cite{steimle2014electric}	&	SnTe	&	2.19	&	2.523	&	259	&	\cite{hoeft1970dipole}	\\
CoO	&	4.18	&	1.621	&		&	\cite{zhuang2014electric}	&	MgO	&	6.2	&	1.749	&	785.1	&	\cite{haynes2014crc}	&	SO	&	1.55	&	1.481	&	1149.2	&	\cite{powell1964microwave}	\\
CrD	&	3.51	&	1.663	&	1182	&	\cite{chen2007permanent}	&	MoC	&	6.07	&		&		&	\cite{wang2007permanent}	&	SrF	&	3.4676	&	2.075	&	502.4	&	\cite{ernst1985electric}	\\
CrN	&	2.31	&	1.5652 \footnotemark[5]	&	854.0 \footnotemark[6]	&	\cite{steimle1999permanent}	&	MoN	&	3.38	&	1.63	&		&	\cite{fletcher1993molecular}	&	SrO	&	8.9	&	1.92	&	653.5	&	\cite{radzig2012reference}	\\
CrO	&	3.88	&	1.615	&	898.4	&	\cite{steimle1989permanentCrO}	&	NaBr	&	9.1183	&	2.502	&	302.1	&	\cite{hebert1968dipole}	&	ThO	&	3.534	&	1.84	&	895.8	&	\cite{wang2011communication}	\\
CS	&	1.958	&	1.535	&	1285.1	&	\cite{winnewisser1968dipole}	&	NaCl	&	9.002	&	2.361	&	366	&	\cite{hebert1968dipole}	&	ThS	&	4.58	&	2.35	&	477  \footnotemark[7]	&	\cite{le2014permanent}	\\
CsBr	&	10.82	&	3.072	&	149.7	&	\cite{story1976dipole}	&	NaCs	&	4.7	&	3.851	&	98.9	&	\cite{radzig2012reference}	&	TiH	&	2.455	&		&		&	\cite{steimle1991laser}	\\
CsCl	&	10.387	&	2.906	&	214.2	&	\cite{hebert1968dipole}	&	NaF	&	8.1558	&	1.926	&	536	&	\cite{hollowell1964radio}	&	TiO	&	3.34	&	1.62	&	1009	&	\cite{steimle2003permanent}	\\
CSe	&	1.99	&	1.676	&	1035.4	&	\cite{mcgurk1973detection}	&	NaH	&	6.4	&	1.889	&	1176	&	\cite{dagdigian1979ground}	&	TiN	&	3.56	&	1.582  \footnotemark[8]	&	1039  \footnotemark[9]	&	\cite{Simard19907012}	\\
CsF	&	7.8839	&	2.345	&	352.6	&	\cite{hebert1968dipole}	&	NaI	&	9.2357	&	2.711	&	258	&	\cite{hebert1968dipole}	&	TlBr	&	4.49	&	2.618	&	192.1	&	\cite{radzig2012reference}	\\
CsI	&	11.69	&	3.315	&	119.2	&	\cite{story1976dipole}	&	NaK	&	2.693	&	3.589	&	124.1	&	\cite{haynes2014crc}	&	TlCl	&	4.5429	&	2.485	&	283.8	&	\cite{hoeft1970dipole}	\\
CuF	&	5.26	&	1.745	&	622.7	&	\cite{wang2010hyperfine}	&	NaRb	&	3.1	&	3.644	&	106.9	&	\cite{haynes2014crc}	&	TlF	&	4.2282	&	2.084	&	477.3	&	\cite{boeckh1964abhangigkeit}	\\
CuO	&	4.57	&	1.724	&	640.2	&	\cite{zhuang2010permanent}	&	NbN	&	3.26	&	1.663	&		&	\cite{fletcher1993permanent}	&	TlI	&	4.61	&	2.814	&	143	&	\cite{haynes2014crc}	\\
CuS	&	4.31	&	2.051	&	415	&	\cite{steimle1988electronic}	&	NH	&	1.39	&	1.036	&	3282.3	&	\cite{haynes2014crc}	&	VN	&	3.07	&	1.566  \footnotemark[10]	&	1033  \footnotemark[11]	&	\cite{steimle1999permanent}	\\
FeC	&	2.36	&	1.61	&		&	\cite{steimle2002permanent}	&	NiH	&	2.4	&	1.476	&	1926.6	&	\cite{gray1985electric}	&	VO	&	3.355	&	1.592  \footnotemark[12]	&	1011.3	&	\cite{suenram1991microwave}	\\
FeH	&	2.63	&		&		&	\cite{steimle2006molecular}	&	NO	&	0.157	&	1.151	&	1904.2	&	\cite{hoy1975stark}	&	VS	&	5.16	&	2.06	&		&	\cite{zhuang2010permanentVS}	\\
FeO	&	4.7	&	1.6	&	970	&	\cite{steimle1989permanentFeO}	&	NS	&	1.86	&	1.494	&	1218.7	&	\cite{byfleet1971electric}	&	WC	&	3.9	&		&		&	\cite{wang2011communication2}	\\
GaF	&	2.4	&	1.774	&	622.2	&	\cite{radzig2012reference}	&	OD	&	1.653	&	0.97	&	2720.2	&	\cite{radzig2012reference}	&	WN	&	3.77	&	1.67  \footnotemark[13]	&		&	\cite{steimle2004permanent}	\\
GaBr	&	2.45	&	2.352	&	263	&	\cite{hoeft1970dipole}	&	OF	&	0.0043	&	1.354	&	1028.7	&	\cite{haynes2014crc}	&	YbF	&	3.91	&	2.016	&	501.9	&	\cite{sauer1996laser}	\\
GeO	&	3.2824	&	1.625	&	985.5	&	\cite{raymonda1970electric}	&	OH	&	1.6498	&	0.97	&	3737.8	&	\cite{nelson1989dipole}	&	YF	&	1.82	&	1.926	&	631.3	&	\cite{shirley1990molecularYF}	\\
GeS	&	2	&	2.012	&	575.8	&	\cite{hoeft1970dipole}	&	PbO	&	4.64	&	1.922	&	721	&	\cite{hoeft1969elektrisches}	&	YO	&	4.524	&	1.79	&	861	&	\cite{suenram1990pulsed}	\\
GeSe	&	1.648	&	2.135	&	408.7	&	\cite{hoeft1970elektrisches}	&	PbS	&	3.59	&	2.287	&	429.4	&	\cite{hoeft1969elektrisches}	&	ZrO	&	2.551	&	1.712	&	969.8	&	\cite{suenram1990pulsed}	\\

\end{tabular}
\end{ruledtabular}

\begin{minipage}[t]{0.24\linewidth}
\footnotetext[1] {From Ref.\cite{schwerdtfeger1995spectroscopic}. }
\footnotetext[2] {From Ref. \cite{okabayashi2005pure}. }
\footnotetext[3] {From Ref.~\cite{parsons2018high}. }
\footnotetext[4] {From Ref.~\cite{bernath1981laser}. }
\end{minipage}\hfill
\begin{minipage}[t]{0.24\linewidth}

\footnotetext[5] {From Ref.~\cite{sheridan2002rotational}. }
\footnotetext[6] {From Ref.~\cite{harrison1996electronic}. }
\footnotetext[7] {From Ref.~\cite{bartlett2013spectroscopic}. }
\footnotetext[8] {From Ref.~\cite{dunn1970rotational}. }
\end{minipage}\hfill
\begin{minipage}[t]{0.24\linewidth}
\footnotetext[9] {From Ref.~\cite{douglas1980electronic}. }
\footnotetext[10] {From Ref.~\cite{balfour1993rotational}. }
\footnotetext[11] {From Ref.~\cite{simard1989spectroscopy}. }
\footnotetext[12] {From Ref.~\cite{merer1989spectroscopy}. }
\end{minipage}\hfill
\begin{minipage}[t]{0.24\linewidth}
\footnotetext[13] {From Ref.~\cite{ram1994high}. }
\end{minipage}

\end{table*}

\nocite{*}

\bibliography{aipsamp}

\end{document}